\documentclass{superfri}
\usepackage[hidelinks]{hyperref}
\begin{document}

\author{Saurabh Hukerikar and Christian Engelmann
        \footnote{Oak Ridge National Laboratory\\
                  One Bethel Valley Road,\\
                  Oak Ridge, TN 37831 USA\\
		  \vspace{2pt}\\
This manuscript has been authored by UT-Battelle, LLC under Contract No. DE-AC05-00OR22725 with the U.S. Department of Energy. The United States Government retains and the publisher, by accepting the article for publication, acknowledges that the United States Government retains a non-exclusive, paid-up, irrevocable, worldwide license to publish or reproduce the published form of this manuscript, or allow others to do so, for United States Government purposes. The Department of Energy will provide public access to these results of federally sponsored research in accordance with the DOE Public Access Plan (http://energy.gov/downloads/doe-public-access-plan)}
} 

\title{Resilience Design Patterns: A Structured Approach to Resilience at Extreme Scale}

\maketitle{}

\begin{abstract}%
Reliability is a serious concern for future extreme-scale high-performance computing (HPC) systems. Projections based on the current generation of HPC systems and technology roadmaps suggest the prevalence of very high fault rates in future systems. While the HPC community has developed various resilience solutions, application-level techniques as well as system-based solutions, the solution space remains fragmented. There are no formal methods and metrics to integrate the various HPC resilience techniques into composite solutions, nor are there methods to holistically evaluate the adequacy and efficacy of such solutions in terms of their protection coverage, and their performance \& power efficiency characteristics. Additionally, few of the current approaches are portable to newer architectures and software environments that will be deployed on future systems. In this paper, we develop a structured approach to the design, evaluation and optimization of HPC resilience using the concept of design patterns. A design pattern is a general repeatable solution to a commonly occurring problem. We identify the problems caused by various types of faults, errors and failures in HPC systems and the techniques used to deal with these events. Each well-known solution that addresses a specific HPC resilience challenge is described in the form of a pattern. We develop a complete catalog of such resilience design patterns, which may be used by system architects, system software and tools developers, application programmers, as well as users and operators as essential building blocks when designing and deploying resilience solutions. We also develop a design framework that enhances a designer's understanding the opportunities for integrating multiple patterns across layers of the system stack and the important constraints during implementation of the individual patterns. It is also useful for defining mechanisms and interfaces to coordinate flexible fault management across hardware and software components. The resilience patterns and the design framework also enable exploration and evaluation of design alternatives and support optimization of the cost-benefit trade-offs among performance, protection coverage, and power consumption of resilience solutions. The overall goal of this work is to establish a systematic methodology for the design and evaluation of resilience technologies in extreme-scale HPC systems that keep scientific applications running to a correct solution in a timely and cost-efficient manner despite frequent faults, errors, and failures of various types.

\end{abstract}

\section{Introduction}
\label{section:Introduction}

Extreme-scale, high-performance computing (HPC) will significantly advance discovery in fundamental scientific research by enabling multiscale simulations that range from the very small, on quantum and atomic scales, to the very large, on planetary and cosmological scales. Computing at scales in the hundreds of petaflops, exaflops and beyond will also provide the computing power for rapid design and prototyping and big data analysis. Yet, to build and effectively operate extreme-scale HPC systems, there are several key challenges, including management of power, massive concurrency, and resilience \cite{Dongarra:2011:IES}. 

In the pursuit of greater computational capabilities, the architectures of future HPC systems are expected to change radically. These innovative systems require equally novel components, which are designed to communicate and compute at unprecedented rates. Traditional HPC system design methodologies have not had to account for power constraints, or parallelism on the level designers must contemplate for future extreme-scale systems \cite{Shalf:2011}. The evolution in the architectures will require changes to the programming models and the software environment to ensure application scalability. In the midst of these rapid changes, the resilience to faults or defects in system components, which can cause errors and failures, will be critical. The reliability of these systems will be threatened by a decrease in individual transistor reliability due to manufacturing defects prevalent at deeply scaled technology nodes, device aging related effects, etc. \cite{Borkar:2005}. The chips built using these devices will be increasingly susceptible to errors due to the reduced noise margins arising from near-threshold voltage (NTV) operation \cite{Dreslinski:2010} (that will be necessary to meet the limits on power consumption). These effects are expected to increase the rate of transient and hard errors in the system. The scientific applications running on these systems will no longer be able to assume correct behavior of the underlying machine. The errors will propagate and generate various kinds of failures, which may result in outcomes in HPC applications ranging from data corruptions to catastrophic crashes. 

Managing the resilience of future extreme-scale systems is a complex, multidimensional challenge. As HPC systems approach exaflops scale, the sheer frequency of faults and errors in these systems will render many of the existing resilience solutions ineffective. Newer modes of failures due to faults and errors, which will only emerge in advanced process technologies and complex system architectures, will require novel resilience solutions. To remain viable the adaptations of existing solutions, as well as the designs of new solutions, must also navigate the complexity of the hardware and software environments of future systems. Additionally, HPC resilience solutions, both hardware and software, must optimize for some combination of performance, power consumption and cost while providing effective protection against faults, errors and failures. Therefore, addressing the resilience challenge for extreme-scale HPC systems will require integration and coordination between various hardware and software technologies that are collectively capable of handling a broad set of fault models at accelerated fault rates. 

The HPC community and vendors have developed a number of hardware and software resilience solutions over the years to confront faults and their consequences in a HPC system and to limit their impact on the applications. Most of these solutions are based on a limited set of underlying detection, containment and mitigation techniques that have persisted through generations of systems and will remain important in the future. The key to the design and implementation of HPC resilience solutions is no longer the invention of novel methodologies for dealing with the various fault types that may occur, or to manage the extreme fault rates; rather, it is based on the selection and combination of the most appropriate solutions among the well-understood resilience techniques and adapting them to the design concerns and constraints of the emerging extreme-scale systems. However, there are no systematized methods to adapt the existing solutions to future architectures and software environments, nor are there formalized to integrate multiple solutions into composite solutions. There is also a lack of standardized methods to investigate and evaluate the effectiveness and efficiency of such solutions. Therefore, the designers of HPC hardware and software components have a compelling need for a systematic methodology for designing, assessing and optimizing resilience solutions. 

In this work, we develop a structured approach for constructing resilience solutions for HPC systems and their applications based on the concept of design patterns. Design patterns are descriptions of well-known solutions to specific, repeatedly occurring problems that are encountered in a specific domain. In an effort to develop resilience design patterns we identify well-known techniques to handle faults and their consequences in various hardware and software components throughout the HPC system stack. In general, resilience solutions provide techniques for the detection of faults, errors or failures in a system, mechanisms to ensure that their propagation is limited, and for masking of error or failure and recovery of the system. This paper presents a complete catalog of patterns that capture the solutions for each of these three aspects. Each pattern provides a solution to a recurring HPC resilience problem under a set of clearly defined assumptions about the type of the fault, error or failure it deals with and the constraints about the system behavior it guarantees. The resilience design patterns are specified at a high level of abstraction and describe solutions that are free of implementation details. The patterns have the potential to shape the design of HPC applications' algorithms, numerical libraries, system software, and hardware architectures, as well as the interfaces between layers of system abstraction. Therefore, they are intended to be useful for HPC application, library and tool developers, hardware architects and system software designers, as well as system users and operators.  
 
We codify the resilience design patterns in a layered hierarchy, which classifies the patterns in the catalog, and clearly conveys the relationships among them. The hierarchical scheme enables individual hardware/software component designers to focus on problems and constraints related to detection, containment and mitigation/recovery of specific fault types in specific contexts, while system architects contemplate role of the individual patterns within the context of the overall system architecture and software environment and issues related to stitching the various patterns together and refinement of their interactions. Combining these patterns according to the guidelines given by the classification scheme provides a systematic way to design and implement new resilience solutions, port existing solutions to future architectures and software environments, and to holistically evaluate the scope and efficiency of the solutions. Therefore, using the design patterns as building blocks enables:
\begin{itemize}
\item Systematic design and refinement of resilience solutions by using patterns to outline the overall structure of the solution (independent of a specific implementation approach), and incrementally converging towards a detailed implementation. 
\item Design of solutions with a clear understanding of their protection coverage and performance efficiency.
\item Evaluation and comparison of alternative resilience solutions through qualitative and quantitative evaluation of the coverage and handling efficiency of each solution.
\item Design of flexible solutions through integration of multiple patterns into complete resilience solutions. The individual patterns may be independently evolved and developed for portability to different HPC system architectures and software environments.  
\item Design of cross-layered resilience solutions that combine capabilities from different layers of the system stack.
\item Optimization of the trade-off space, at design time or at runtime, between the key system design factors: performance, resilience, and power consumption.
\end{itemize}

In this paper, we also develop a systematic methodology to combine an essential set of patterns into productive and efficient resilience solutions. We present a conceptual framework based on the notion of \textit{design spaces} that enables HPC designers to use the patterns as reusable design elements. The framework enables designers to navigate the complexities of composing patterns into complete solutions within the constraints of performance and power overheads, the fault model and its impact on the system, hardware and software implementation challenges, etc. The overall goal of this work is to enable a systematic methodology for the design and evaluation of resilience technologies in HPC systems that keep applications running to a correct solution in a timely and cost-efficient manner despite frequent faults, errors, and failures of various types.

\section{Design Patterns for HPC Resilience}
\label{sec:Motivation:Concept}

The occurrences of various types of faults, errors and failures are not rare events in modern large-scale HPC system environments. The term \textbf{fault} refers to an underlying flaw or defect in a system that has potential to cause problems, an \textbf{error} refers to the result of the activation of a fault, which causes an illegal system state. A \textbf{failure} occurs if an error reaches the service interface of a system, resulting in system behavior that is inconsistent with the system's specification. The faults are due to radiation-induced effects such as particle strikes from cosmic radiation and the system environment, chip manufacturing defects and design bugs that remain undetected during post-silicon validation and manifest themselves during system operation, as well as circuit wear out, or aging failure mechanisms of CMOS integrated circuits. The faults may also occur due to software bugs, which is a growing concern as the complexity of the software environment grows. Due to the complex system interactions and dependencies between the hardware and software components, the application program, and the HPC system's physical environment, preventing the activation of these faults and containing the propagation of the resulting errors and failures to other components a significant challenge. 

HPC resilience solutions seek effective and efficient management of the different types of fault and errors to ensure that the applications produce reliable outcomes despite the resulting degradations and failures. The focus of resilience solutions is on application correctness lieu of, or even at the expense of, reliability of state of the system. In general, every HPC resilience solution consists of the following core capabilities:
\begin{itemize}
\item \textbf{Detection:}
Identifying the presence of an anomaly in the data or control value is an important aspect of any resilience management strategy. The detection and diagnosis of faults in a system may allow the remedy of the underlying defect, which may prevent the activation of an error or failure. The timely detection of errors or failures enables recovery of the system. 

\item \textbf{Containment:}
When an error or failure is discovered in a system, containment strategies assist in limiting the impact of the event on other components in the system. Limiting the propagation enables simplified recovery strategies. 

\item \textbf{Recovery:}
The recovery aspect of any resilience solution is necessary to ensure that the application outcome is correct in spite of the presence of an error or a failure in a system. The recovery may entail a workaround to isolate and bypass the presence of an error or a failed component, complete elimination of the error or failure, and may also seek to prevent the root cause of the underlying fault from resurfacing. 
\end{itemize}

Often the solutions used to achieve these capabilities are based on well-known techniques, which have been repeatedly used by hardware and software designers to increase system reliability since the early days of computing systems. These techniques are based on the use of redundant structures to mask failed components, error-control codes and duplication or triplication with voting to detect or correct information errors, diagnostic techniques to locate failed components, automatic switchovers to replace failed subsystems, and the specification of well-defined modular structures and interfaces for containment and definition of recovery scope \cite{Avizienis:1997}. Many of the resilience solutions, hardware and software, used in HPC environments over the past three decades are also largely based on these set of techniques. 

Our goal is to capture the best-known techniques that are used in the design of HPC resilience solutions formatted as design patterns. A design pattern describes a generalizable solution to a recurring problem that occurs within a well-defined context. It identifies the key aspects of a solution and presents it in the form of an abstract description, which provides designers with guidelines on how to solve a problem. Each pattern in this paper presents a solution to a specific problem in detecting, recovering from, or masking a fault, error or failure event. The pattern descriptions don't describe a concrete design or an implementation, and are also free from constraints of details associated with the level of system abstraction at which the solution can be implemented. Therefore, the resilience patterns may be used as design templates that may be adapted by the HPC hardware or software designers for a specific problem at hand.  
The design of new resilience solutions and adapting existing ones for future extreme-scale systems is accomplished by combining various patterns into complete solutions and by refining their interactions. The patterns describe the design decisions and trade-offs that must be considered when applying a certain solution, which enables designers to reason about the impact of applying a solution on a system's performance scalability and power consumption overhead as well as consider implementation issues. Since the various resilience techniques handle different types of events, and they each provide different guarantees about properties such as the time or the space overhead introduced to the normal execution of the system, number of simultaneous errors or failures it can handle, the efficiency of the reaction to a failure, the design complexity added to the system, the patterns may also be used to explore alternatives solutions to a given problem. 

Based on the insight that any resilience solution is only necessary in the presence of, or sometimes in the anticipation of an anomalous event, such as a fault, error, or failure, we define the template of a resilience design pattern in an event-driven paradigm. The design pattern template consists of a {\em behavior} and a set of {\em activation} and {\em response interfaces}. The pattern behavior provides a description of the solution, which systematically names, explains the semantics of, and evaluates the trade-offs involved in using the solution in an HPC environment. The activation and response interfaces specify the conditions for application of the solution. The individual implementations of the same pattern may have different levels of performance, resilience, and power consumption. However, using this universal template enables a standardized approach for the evaluation of patterns and comparison between alternative solutions for a given problem. While any resilience design pattern must conform to this basic template, the instantiation of a pattern may cover combinations of detection, containment and mitigation capabilities.

\section{Classification of Resilience Design Patterns}
\label{sec:RDP-Classification}
\begin{figure*}[tp]
  \centering
  \includegraphics[width=\textwidth]
	          {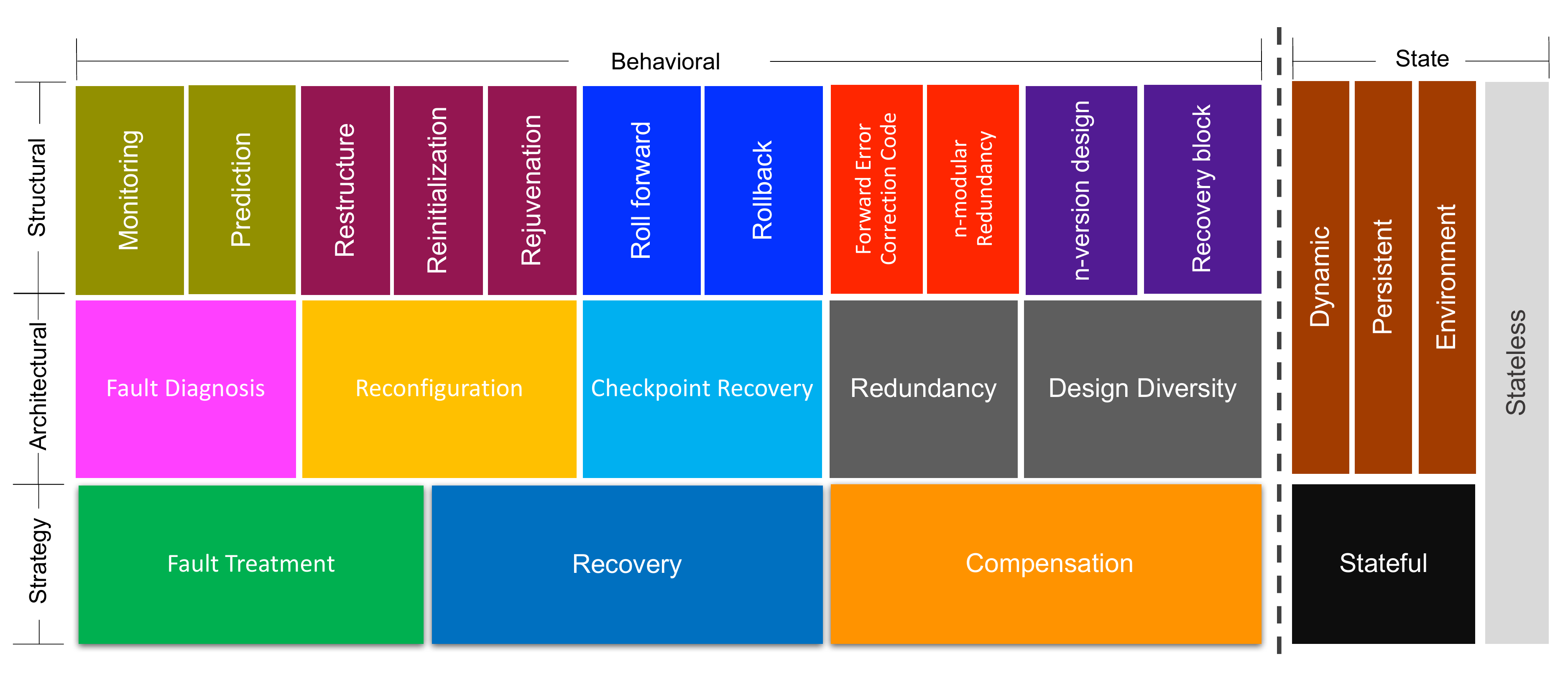}
  \caption{Classification of resilience design patterns}
  \label{Fig:PatternClassification}
\end{figure*}
For designers of HPC resilience solutions, the patterns serve as reusable design elements. For the design of resilient hardware and software components, the patterns can be combined in different ways to produce complete solutions. For a systematic approach to transforming individual patterns into a solution consisting of a system of patterns, a classification scheme is essential. A classification outlines the relationships between the various patterns, which enables designers to understand their individual capabilities and the relationships among the patterns when seeking to integrate different patterns into composite solutions. 

The resilience design patterns may be classified on the basis of the type of event handled, whether the pattern offers detection, containment, recovery or masking semantics, the scope of protection coverage offered, design complexity of the patterns, time and space overheads, power consumption overheads, etc. However, in developing a classification for the resilience design patterns, our goal is to provide designers with the guidelines to identify the patterns that make up a resilience solution, specify the roles played by individual patterns, and how they interact, such that the incorporation of resilience capabilities becomes an essential part of the design process of HPC hardware and software components. 

We develop a pattern classification scheme that organizes the resilience patterns in a layered hierarchy, in which each level addresses a specific aspect of the problem. Resilience in the context of HPC systems and its applications has two key dimensions: (1) forward progress of the system and (2) data consistency in the system. Based on these factors, we organize the resilience design patterns into two major categories, \textbf{state} patterns and \textbf{behavioral} patterns. These are placed side by side in Figure \ref{Fig:PatternClassification} to enable designers to separately reason about the patterns that define the scope of the protection domain and those that define the semantics of the detection, containment and mitigation. The behavioral patterns are organized in a hierarchy, as shown in Figure \ref{Fig:PatternClassification}, in which the patterns in bottom layer may be used to think about the strategies suitable for confronting anomalous events depending on whether it is a fault, an error, or a failure. The patterns in the middle layer explicitly defines the architecture of a solution based on the nature of the event and considers compatibility of the pattern solution with the overall system design. The top-level patterns consider issues related to implementation of the solution, including the appropriate granularity and level of system abstraction, and the overheads incurred by the solution. 

For the design and analysis of new solutions, or adapting existing solutions to emerging HPC environments, hardware and software designers can approach the hierarchy of patterns in a top down or bottom-up manner. The refinement and optimization of patterns will often require traversing the layers several times before a solution is finalized. The hierarchical organization of the patterns permits the different stakeholders to reason about resilience solutions based on their view of the system and their core expertise. Architects describe the overall organization of the solutions, analyze the integration of various resilience patterns across the system stack, and evaluate the protection coverage and overheads to overall system performance. The designers of individual hardware and software components operate within a single layer of system abstraction and focus on alternative patterns to address the problem at hand and the analyze the design complexity of instantiating a specific pattern.   

\subsection{State Patterns}
The state patterns specify the protection domain of a resilience solution. The correctness and consistency of the system state ensures the correct operation of a system. Therefore, the precise definition of the scope of the protected system state is an important part of designing a resilience solution. From the perspective of an HPC application, the notion of state may be classified into three categories:
\begin{itemize}
\item \textit{Static State}, which represents the data that is computed once in the initialization phase of the application and is unchanged thereafter.
\item \textit{Dynamic State}, which includes all the system state whose value may change during the computation.
\item \textit{Operating Environment State}, which includes the data needed to perform the computation, i.e., the program code, environment variables, libraries, etc.
\end{itemize}

The state patterns, which capture each of these aspects of the system state, are classified as \emph{stateful} patterns. The properties of each state pattern may be used to guide the selection of a behavioral pattern. Certain resilience strategies may be applied without regard for state and apply behavioral patterns that are concerned with only the forward progress of the system (for e.g., idempotent operations). Therefore, the classification of state patterns also includes a \emph{stateless} pattern that enable designers to create solutions that define behavior without state. This organization of the state patterns enables the behavioral patterns to be applied to individual aspects of a system's state. However, in designing a resilience solution, more than one type of state pattern may be fused to enable the use of a single behavioral pattern for more than one state pattern.  

\subsection{Behavioral Patterns}
The behavioral patterns are concerned with forward progress of the system despite the presence of anomalous events in the system. These design patterns identify detection, containment, or mitigation actions that enable the components in a system that realize these patterns to cope with the presence of a fault, error, or failure event. The behavioral patterns are presented in a layered hierarchy to highlight the design choices when selecting one pattern over another: 
\begin{itemize}
\item \textbf{Strategy Patterns:}
These patterns define high-level polices of a resilience solution. The strategy patterns are organized by the type of event that they handle - fault, error or failure, since the techniques to handle these events are fundamentally different. The classification of the strategy patterns captures the intent behind of each solution makes the design choices in applying the patterns explicit. These patterns describe the overall structure and the key components in a solution in a manner independent of the layer of system stack and hardware/software architectural features. Their descriptions are deliberately abstract to enable hardware and software architects to reason about the overall organization of the solution and assess the suitability of the pattern to the full system design. 

The fault treatment patterns are concerned with diagnosing and preventing an imminent error or failure. The recovery and compensation patterns must limit and remove an error or failure state in the system. The recovery pattern aims to substitute an error/failure-free state in place of the erroneous/failed system state. The compensation pattern seeks to tolerate the presence of an error or failure by providing redundancy in the system design.

\item \textbf{Architectural Patterns:}
The architectural patterns convey specific methods necessary for the construction of a resilience solution. The patterns provide details about the key components and connectors that make up the solution and explicitly specify the type of event that they handle. These patterns are a sub-class of the strategy patterns, and they are also organized based on the type of event they handle and the intended impact of the action on the system resilience. Certain architectural patterns may be adapted to confront faults, errors or failures. Consequently, there exists an overlap between the patterns in the architectural layer with more than one type of strategy pattern in Figure \ref{Fig:PatternClassification}. The classification of these architectural patterns based on the core solution is also suggestive of the design time and runtime complexity encountered when instantiating a pattern. Yet, architectural pattern descriptions are independent of the precise fault model and may be implemented at any layer of the system stack.

\item \textbf{Structural Patterns:}
These patterns provide concrete descriptions of the solution rather than high-level strategies. While the strategy and architectural patterns serve to provide designers with a clear overall framework of a solution and the type of events that it can handle, the structural patterns express the details such that they can contribute to the development of complete working solutions. They comprise of specific instructions for implementing the pattern, including concrete descriptions of the key parts of the solution. Their descriptions include specific details of the fault model that the pattern handles. Although the structural patterns provide more detailed solutions, their descriptions are still independent of the layer of system abstraction at which the patterns may be instantiated. The pattern descriptions are flexible enough for most, if not all structural patterns to be suitable for implementation within hardware structures as well as within algorithms in the application or system software. The various structural patterns are sub-classes of the strategy and architectural patterns. Therefore, their first-order organization is also based on the type of fault event that their solutions handle. 
\end{itemize}

A variety of {\bf implementation patterns} may be derived from the structural patterns. These patterns are intended to bridge the gap between the design principles and the concrete details of an implementation. The pattern descriptions explicitly specify the layer of system abstraction at which they are implemented, and the activation and response interfaces. The implementation patterns also enable a standardized way for hardware and software designers to communicate about design of their resilience solutions. These patterns may be designed as composite patterns, i.e., using a combination of patterns. Defining implementation patterns enables designers to thoroughly analyze the overhead of a solution in terms of time and space, as well as the trade-off between design complexity and runtime complexity. Due to the limitless possibilities in developing implementation patterns suited for various architectures, software environments and HPC applications through pattern composition, we only provide detailed descriptions of the foundational state and behavioral resilience patterns in this paper.

\section{The Resilience Design Pattern Catalog}

The resilience design pattern catalog contains detailed descriptions of the state and behavioral patterns. The primary objective of the catalog is to capture the best-known HPC resilience solutions and present them a standardized and accessible form. For the patterns to be useful to HPC system architects and individual hardware and software component designers alike, they are written down in a highly structured format to enable designers to quickly discover whether the pattern solution is suitable to the problem being solved. 

For convenience and clarity, each resilience pattern in the catalog follows the same prescribed format. The pattern description is formatted in terms of the following key attributes: 
\begin{itemize}
\item\textbf{\emph{Name}}:
Identifies the pattern and provides a convenient way to refer to it, typically using a short phrase.

\item\textbf{\emph{Problem}}:
A description of the problem indicating the intent behind applying the pattern. This describes the goals and objectives that will accomplished with the use of this specific pattern.

\item\textbf{\emph{Context}}:
The preconditions under which the pattern is relevant, including a description of the system before the pattern is applied.

\item\textbf{\emph{Forces}}:
A description of the relevant forces and constraints, and how they interact or conflict with each other, and with the intended goals and objectives. The forces highlight the intricacies of the problem and make the trade-offs that must be considered explicit.

\item\textbf{\emph{Solution}}:
A description of the solution that includes specifics of how to achieve the intended goals and objectives. This includes the core structure of the solution, its semantics and its interactions with other patterns. The description includes guidelines for implementing the solution, as well as descriptions of variations or specializations of the solution.

\item\textbf{\emph{Capability}}:
The resilience management capabilities provided by this pattern, which may include detection, containment, mitigation, or a combination of these capabilities. 

\item\textbf{\emph{Protection Domain}}:
The resiliency behavior provided by the pattern extends over a certain scope, which may not always be explicit. The description of the nature of the fault model and its protection domain enables designers to reason about the scope of the coverage in terms of the complete system.

\item\textbf{\emph{Resulting Context}}:
A brief description of the post-conditions arising from the application of the pattern. There may be trade-offs between competing optimization parameters that arise due to the use of this pattern. 

\item\textbf{\emph{Examples}}:
One or more sample applications of the pattern, which illustrate the use of the pattern for a specific problem, the context, and set of forces. This also includes a description of how the pattern is applied, and the resulting context.

\item\textbf{\emph{Rationale}}:
An explanation of the pattern as a whole with an elaborate description of how the pattern actually works for specific situations. This provides insight into its internal workings of a resilience pattern, including details on how the pattern accomplishes the intended goals.

\item\textbf{\emph{Related Patterns}}:
The relationships between this pattern and other relevant patterns. These patterns may be predecessor or successor patterns in the hierarchical classification, or patterns that provide similar capabilities. 

\item\textbf{\emph{Known Uses}}:
Known applications of the pattern in existing HPC systems, including any practical considerations and limitations that arise due to the use of the pattern at scale in production HPC environments.

\end{itemize}

There are three key reasons behind this pattern format: (1) to present the pattern solution in a manner that simplifies comparison of the capabilities of patterns and their use in developing complete resilience solutions, (2) to present the solution in a sufficiently abstract manner that designers may modify the solution depending on the context and other optimization parameters, and (3) to enable these patterns to be instantiated at different layers in the system. 

The complete catalog of resilience design patterns in the template format is available in a specification document \cite{RDP:Website}. In the remainder of this section, we summarize each design pattern, highlighting its key features. The pattern descriptions use the term \textit{system} to refer to an entity that has the notion of a well-defined structure and behavior. A \textit{subsystem} is a set of elements, which is a system itself, and is a component of a larger system, i.e., a system is composed of multiple sub-systems or components. For a HPC system architect, the scope of system may include compute nodes, I/O nodes, network interfaces, disks, etc., while an application developer may refer to a library interface, a function, or even a single variable as a system. A \textit{full system} refers to the HPC system as a whole or to a collection of nodes, which is capable of running a parallel application.

%
%
\subsection{Strategy Patterns}
\subsubsection{Fault Treatment Pattern\\}
The emergence of a defect or anomaly in an HPC system environment has the potential to activate, which may potentially lead to an error or a failure in the system. The \texttt{Fault Treatment} pattern provides a method that attempts to recognize the presence of an anomaly or a defect within a system, and creates conditions that prevents the activation of the fault into an error or failed state. The solution requires an auxiliary monitoring system, which may be a sub-system of the monitored system or an external system, that observes the key parameters of the monitored system. The pattern applies to a system that has well-defined parameters that may be used to discover the presence of anomalies in the behavior of the monitored system. The pattern supports either one, or both of the following capabilities:
\begin{itemize}
\item \textbf{Fault detection}: detect anomalies during operation before they impact the correctness of the system state. 
\item \textbf{Fault mitigation}: methods to enable an imminent error or failure to be prevented, or the defect to be removed.
\end{itemize}

The protection domain of this pattern extends to the scope of the monitored system and implicitly extends to other systems that are interfaced to the monitored system. The benefit of incorporating fault treatment patterns in a design, or deploying it during system operation is to preemptively recognize faults in the system; the preventive actions avoid the need for expensive recovery and/or compensation actions that may be necessary if the fault activation causes an error or failure. In incorporating this pattern in the design of a HPC hardware or software component, the key considerations are the frequency of interaction between the monitoring and monitored (sub-)systems and the precision of fault detection. The frequency of these interactions must be minimized to reduce interference in the operation of the monitored system; yet, the interactions must be frequent enough to detect every defect in the monitored system. Also, fault must be detected and treated in a timely manner, i.e., the time interval for the monitoring system to gather data about the monitored system and to infer the presence of an anomaly or a defect must be rapid to prevent the activation of an error/failure. The pattern must also have few false positive and false negatives to minimize preemptive mitigation actions that are unnecessary.  

In HPC systems, various hardware-based solutions for fault detection observe the attributes of a system, such as thermal state, timing violations in order to determine the presence of a defect in the behavior of the system that may potentially cause an error or failure. For example, processor chips such as the IBM Power 8 and Intel Xeon series processors contain thermal sensors that detect anomalous conditions in the cores. Software-based solutions detect the anomalies in the behavior of a system's data variables or control flow attributes to determine the presence of a fault. Heartbeat monitoring is used for liveness checking of MPI processes, which enables detection of imminent failure of the MPI communicator \cite{Batchu:2004}. \newpage 

\subsubsection{Recovery Pattern\\}
In an HPC environment, the occurrence of errors or failures in the system results in incorrect answers, and in some cases, catastrophic application crashes. The \texttt{Recovery} pattern enables a system to survive an error or failure event. The pattern is suitable for a system whose design or runtime configuration contains no intrinsic support for tolerating the error or failure event. The solution is based on the periodic creation of snapshots of the system state during error/failure-free operation, and the maintenance of these snapshots persistently. Upon detection of an error or a failure, the preserved snapshots are used to recreate a known error/failure-free state of the system. When the system state is recovered, the operation of the system is resumed. The error or failure in the system must be detected; the pattern offers no implicit error/failure detection. The pattern applies to a system that is deterministic, i.e. forward progress of the system is defined in terms of the input state to the system and the execution steps completed since system initialization. The pattern requires the system state can be compartmentalized in a form that is accurately representative of the progress of the system since initialization. It also requires that the system has well-defined intervals that enables it to transition the system state to a known correct interval in response to an error/failure. 

The protection domain for a \texttt{Recovery} pattern is determined by the scope of the state pattern that is captured during checkpoint creation operation. The size and frequency of creation of checkpoints determines the overhead to system operation; frequent checkpointing incurs proportionally greater overheads during error/failure-free operation, but reduces the amount of lost work when an error/failure event does occur. Also, the broader the scope of the system state that is preserved, the larger is the scope of the system state that may be protected from an error/failure event. The solution offered by this pattern is not dependent on the precise semantics of the error/failure propagation. Therefore, the effort and complexity in using this pattern in a hardware or software design, or in the system configuration is low. There are several instances of the usage of the pattern in HPC systems to support recovery of an application or the complete system upon detection of an error/failure. For example, various checkpoint and rollback protocols enable HPC applications and systems to capture state and commit the checkpoint files to parallel file systems \cite{Elnozahy:2002}. \\ 

\subsubsection{Compensation Pattern\\}
The occurrence of an error or failure event may cause loss in system functionality, or reduction in system capacity. The HPC applications running on such a system may produce incorrect results or experience failure. The \texttt{Compensation} pattern makes up for the deficiency or abnormality in a system that is caused by the error or failure event. The pattern solution introduces redundancy with into the system design, or in the configuration to counterbalance the (sub-)systems in error or failed state. The pattern is applicable to a system that is deterministic, and the overall system design allows for modular design with well-defined inputs and outputs for each module, about which redundant information is maintained. The redundancy may be in the form of a group of replicas of a (sub-)system, referred to as n-modular redundancy, or in the form of encoded information about the (sub-)system state. The pattern supports detection, and in some cases correction, by using the redundant information about a (sub-) system to recompense for the presence of an error/failure. The scope of the protection domain, which covers includes the part of the system designed or operated redundantly, may include a sub-system, or the cover the full system. 

The replicas of the modules permit the system to continue operation even in the presence of a (sub-)system failure. When the redundancy is in the form of modular replication, an error or failure in one of the (sub-)systems may be tolerated by substituting the (sub-)system with a replica. In order to recover from 2N errors/failures in the system, there must be 2N + 1 distinct replicas. For the detection of errors, the outputs of the replicas of the system are compared by an auxiliary monitor (sub)-system. For a system to tolerate an error/failure, the number of replicas must be greater than two, in which case the monitor performs majority voting on the outputs produced by the replicas. This enables incorrect outputs from replicas in erroneous state to be filtered out. The design effort and complexity of replication of the system depends on the replication method: deploying identical replicas requires low design effort, but the design of functionally identical but independently designed versions of a (sub-)system requires much higher design and verification effort. 

The scope and strength of the redundancy employed by the pattern determines the overhead to the system performance. The pattern introduces a penalty in terms of time (increase in execution time), or space (increase in resources required) independent of whether an errors or failure occurs during system operation. The N-modular redundancy approach is used at the hardware and software levels in a various HPC components; the dual-modular redundancy (DMR) for error detection and triple-modular redundancy (TMR) for error detection and correction \cite{Koren:1979} are the most widely used forms of redundancy. Redundant information in the form of error correction codes is also used at the hardware-level in the form of ECC \cite{Moon:2005} and at the application-level for application data structures \cite{Huang:1984}. 

%
%
\subsection{Architectural Patterns}
\subsubsection{Fault Diagnosis Pattern\\}
The occurrence of a defect or anomaly has the potential to activate causing an error or failure in the system. The \texttt{Fault Diagnosis} pattern, which is a derivative of the \texttt{Fault Treatment} strategy pattern, identifies the presence of the fault and determines its root cause. The solution consists of an auxiliary monitoring system that observes specific parameters of a monitored system. Until a fault has not activated into an error it does not affect the correct operation of the system. Therefore, the \texttt{Fault Diagnosis} pattern makes an assessment about the presence of a defect based on observed behavior of one or more system parameters. The inference is based on observing deviations in the standard operating behavior of the monitored system. Identifying the norm of (sub-)system parameters also enables narrowing the search for the fault type, its location and its root cause. 
To incorporate this pattern in an HPC environment requires inclusion of a monitoring (sub-)system, which introduces additional complexity in the overall system design. When the monitoring system is extrinsic to the monitored system, the design effort may be simplified, but the interfaces between the (sub-)systems must be well-defined. The pattern only infers the presence of a defect and reports it via its response interface, but does not act to remedy the fault. 
Among the key design challenges when using the pattern is the resolution limit, which is influenced by the number of parameters observed and frequency of probing the monitored (sub-)system and affects the precision of the fault detection. In the context of HPC systems, faults may be detected and diagnosed based by accumulating empirical data on the characteristics and the behavior of hardware and software components and use the information to discover faults. For example, HPC components commonly use the Intelligent Platform Management Interface (IPMI) \cite{IPMI}, which provides standardized interfaces for monitoring hardware health information such as the system temperatures, fans, power supplies, etc. Using these interfaces, software tools may monitor the health of system resources and infer the presence of anomalies in the components.

\subsubsection{Reconfiguration Pattern\\}
In the event of a fault, error or failure event the configuration, i.e., the organization of the (sub-)systems in an HPC environment may be affected in ways that result in applications producing incorrect results, or experiencing fatal crashes. The \texttt{Reconfiguration} pattern, which derives from the \texttt{Fault Treatment} and \texttt{Recovery} strategy patterns, entails modification of the interconnection between (sub)-systems. The reconfiguration isolates the (sub-)system affected by the event to prevent it from affecting the correct operation of the overall system. The pattern assumes that the system may be partitioned into a set of logical modules and that altering the interconnection between the modules is possible. The protection domain of the \texttt{Reconfiguration} pattern covers all (sub-)systems that are interconnected to provide a specified function. The pattern may cause the system to assume several configurations in response to a fault, error or failure event, each of which is characterized by its own topology of interconnections, the system must retain functional equivalency with the original system configuration. The performance overhead of using this pattern is proportional to the number of (sub-)systems and degree of interconnection between them. 
The reconfiguration of the system may also result in system operation at a degraded performance level. The implementation of the pattern requires partitioning the system into modules that remain functionally correct in multiple different configurations. There is much complexity associated with defining the scope of these modules and to validate their functional equivalency in alternative configurations. Well-known use cases of the reconfiguration pattern include the NodeKARE module in the Cray Linux Environment {CLE}, which automatically runs diagnostics on all involved compute nodes in the cluster whenever a user’s program terminates abnormally and removes the failing nodes from the pool of available compute nodes so that subsequent jobs are allocated only to healthy nodes \cite{Cray:XE6Spec:2010}.\\  

\subsubsection{Checkpoint Recovery Pattern\\}
Errors or failures in an HPC environment may result in conditions that prevent forward progress of the system until the error or failure condition is removed. The \texttt{Checkpoint-Recovery} pattern, which is a specialization of the \texttt{Recovery} strategy pattern, is based on the creation of snapshots of the system state and maintenance of these checkpoints on a persistent storage system during the error- or failure-free operation of the system. Upon detection of an error or a failure, the checkpoints/logged events are used to recreate last known error- or failure-free state of the system, after which the system operation is restarted. The solution offered by the pattern supports only recovery; the detection and containment of the error/failure is beyond the scope of the pattern's capabilities. The pattern assumes that the system is capable of compartmentalizing its state in a way that is accurately representative of the progress of the system since initialization. The techniques used by the pattern are classified into checkpoint-based and log-based strategies. The checkpoint-based solution typically captures and preserves the complete state of the system; in contrast, log-based strategies only record specific system events. Instantiations of the pattern may also use a combination of checkpointing and event logging. The pattern handles an error or a failure by retrieving a version of the error or failure-free state from the checkpointed state, and substituting the erroneous or failed state with the error or failure-free state. Therefore, the system is able to resume operation with a version of the system state that is free of any effects of the error or failure event. 

However, the pattern requires interruption of the system during error or failure-free operation to record the checkpoint, which incurs an overhead. The frequency of creation of checkpoints and/or event logging determines the extent of the overhead; frequent checkpointing/logging incurs proportionally greater overheads during error- or failure-free operation. However, more frequent checkpointing and logging reduces the amount of lost work when the system encounters an error or failure event. The checkpointing/logging latency affects the overhead during error- or failure-free operation on account of the latency to write the checkpoint to a storage system. The scope of the system state captured during a checkpointing operation results in a proportionate increase in space overhead due to the storage resources needed to preserve the checkpoints. The solution offered by this pattern is independent of the type of error or failure and its mode of propagation. Therefore, the design effort and complexity in instantiating this pattern in any system design in low. In the context of HPC systems, checkpoint and restart capabilities in the software layers, including various library-based and operating system-based solutions such as BLCR \cite{BLCR:2002:LBNL} for Linux processes. Certain library implementations of the MPI standard, such as OpenMPI, also support transparent checkpoint-restart \cite{Hursey:2009}. Log-based recovery based on message logging has been adopted by implementations of MPI \cite{Bouteiller:2010}. 

\subsubsection{Redundancy Pattern\\}
When an error or failure event in an HPC environment cannot be prevented from affecting the correct operation of a component, or the full system, it must be remedied to enable forward progress of the system. The \texttt{Redundancy} pattern, which is a derivative of the \texttt{Compensation} pattern, enables offsetting the effects of the error/failure. The pattern solution entails incorporating excess resources in the (sub-)system design or in the configuration at runtime. The redundancy enables a (sub-)system to detect, and in certain cases correct an error/failure, by repetition, omission of a (sub-)system without loss of functionality, or superfluity of (sub-)system state information. The pattern applies to a (sub-)system that allows for a modular design with well-defined inputs and outputs for each module. The application of a \texttt{Redundancy} architecture pattern, the following error/failure handling capabilities can be supported:  
\begin{itemize}
\item \textbf{Detection by comparison}: observing the likeness of each replica's outputs as means to detect the presence of an error or failure in each redundant version of a (sub-)system.
\item \textbf{Fail-over mitigation}: substitution of a replica in error or failed state with another identical replica that is error/failure-free.
\item \textbf{Mitigation by isolation}: creation of a group of N replicas of a (sub-)system and majority voting on the outputs produced by each replica; the outputs that fall outside the majority are excluded.  
\item \textbf{Encoding information for detection and mitigation}: maintenance of additional (sub-)system state information to identify errors within the state.  
\end{itemize}
 
The protection domain of the pattern extends to the scope of the (sub-)system state about which redundant information is maintained. The pattern introduces penalty in terms of time (increase in execution time), or space (increase in resources required) independent of whether an errors or failure occurs. The use of dual-modular redundancy for error detection and triple-modular redundancy for error/failure detection and correction are common forms of instantiation of the pattern in various hardware and software-level modules. HPC systems contain service nodes that are responsible for system management tasks while the parallel computation is performed by a set of compute nodes. The tasks include user login, network file system, job and resource management, communication services. Various existing solutions provide hot-standby redundancy with transparent fail-over to tolerate failures in the critical services in the service nodes. Well-known examples of redundancy are the scheduling and resource management services in Simple Linux Utility for Resource Management (SLURM) \cite{Slurm} , as well as the metadata servers of the Parallel Virtual File System (PVFS) \cite{PVFS-HA} and the Lustre file system \cite{Lustre-HA}. Production HPC systems such as the Cray XC40 series \cite{Cray:XC40Spec:2014} include redundant power supplies, voltage regulator modules and cooling fans to ensure continuous operation in the event that one of these units experience malfunction or failure. 

\subsubsection{Design Diversity Pattern\\}
Design flaws on account of human error or defective tools manifest themselves as errors, which may cause failures in HPC environments. The \texttt{Design Diversity} pattern, which is also a derivative of the \texttt{Compensation} pattern,  creates distinct but functionally equivalent versions of the same design specification, which are created by different individuals or teams, or developed using different tools. The intent behind applying this pattern is to eliminate the impact of design bugs during the implementation of a (sub-)system. The pattern enables systems to tolerate errors/failures due to design faults that may arise on account of incorrect interpretation of the specifications by designers, mistakes made during implementation, or due to bugs in the tools. The detection and correction of error/failures is possible due to the independent design processes reducing the likelihood that the same flaw emerges in the alternative versions of a (sub-)system. The pattern is based on the assumption that the system has a well-defined specification for which multiple implementation variants may be created. 
The versions of the (sub-)system specification may be applied to a system in a time or space redundant manner. The replica (sub-)systems are provided with identical inputs, and their respective outputs are compared in order to detect and potentially correct the impact of an error or a failure in either replica of the systems. The protection domain of the pattern extends to the scope of the system that is described by the design specification. However, designing multiple variants of the same (sub-)system specification requires significantly higher verification and validation effort. The design diversity solution is used in the validation of the results produced by scientific applications, particularly those that require high-precision floating point arithmetic. Such applications may be compiled and executed using alternative implementations of compiler toolchains, message passing libraries, numerical analysis libraries to verify the application results. \\ 

%
%
\subsection{Structural Patterns}
\subsubsection{Monitoring Pattern\\}
The various types of errors in HPC environments occur as a result of underlying defects in hardware or software components. Identifying the defects before they cause an error, which may result in a failure of one or more components, prevents incorrect behavior of a (sub-)system. The \texttt{Monitoring Pattern} is a specialization of the \texttt{Fault Diagnosis} architectural pattern, which consists of a monitoring system that observes specific parameters of a monitored system to discover the presence of anomalies in its behavior. The monitoring system may approach the problem of fault detection using two strategies:
\begin{itemize}
\item \textit{Effect-Cause Diagnosis}: This approach entails observation of the parameters of the (sub-)system for anomalies. When a (sub-)system parameter deviates from a range of values considered \textit{normal}, the monitoring system attempts to determine the root cause. The monitoring system logically partitions the system into modules and progressively eliminates the modules known to be fault-free. Through this process, it narrows the search for the fault in the (sub-)system.
\item \textit{Cause-Effect Diagnosis}: This approach is based on a set of known fault models and the monitoring system compares the (sub-)system parameters with a model developed using fault free system operation, or using simulations. When the observed set of parameters deviates from a model, the presence of and the type of fault may be inferred.
\end{itemize}

Based on these inferences, the pattern enables the monitored system to report the presence of a fault and to analyze its root cause and location. The (sub-)system design or configuration must include a monitoring (sub-)system. When the monitoring system is extrinsic to the monitored system, the design effort may be simplified, but the interfaces between the (sub-)systems must be well-defined. However, when the monitoring system is intrinsic to the design or configuration of the monitored system, the complexity of the design process increases. The \texttt{Monitoring} pattern only infers the presence of a defect and reports it, but does not remedy the defect. Various HPC system installations use the monitoring pattern through tools for collecting performance- or health-related data about the system. Popular solutions include: Ganglia Monitoring System \cite{Ganglia}, Nagios \cite{Nagios} and OVIS Lightweight Distributed Monitoring System \cite{LDMS:SC14}.

\subsubsection{Prediction Pattern\\}
The accurate prediction of where faults are likely to occur in a (sub-)system enables reduction in the costs of a resilience solution by preemptively enhancing the (sub-)system's capabilities to handle any resulting errors or failures. The \texttt{Prediction Pattern}, which is also a derivative of the \texttt{Fault Diagnosis} architectural pattern, develops models that estimate future faults based on the observations of the parameters of a (sub-)system, or based historical trend analysis of these parameters. For prediction, the pattern may use: (i) \textit{Rule-based methods} that build rules of association to capture the causal correlations between system parameter values and fault events, or (ii)\textit{Statistical-based methods} that discover probabilistic characteristics of potential errors/failures in a system using statistical inference techniques to examine correlations between previous events. The monitoring system of this pattern contains the following components: 
\begin{itemize}
\item \textit{Filter/Preprocessor}: removes incomplete fault data and duplicates and produces a consistent format for analysis.
\item \textit{Regression}: seeks to analyze the parameter values and establish relationships between them.
\item \textit{Knowledge Base}: storage component that maintains the rules or statistical properties and models, which may be used for online prediction of fault events using real-time data captured from the monitored system.
\end{itemize}

Much like the \texttt{Monitoring Pattern}, the \texttt{Prediction} pattern only infers the presence of a defect and reports it, but does act to remedy the fault. Based on the prediction method and accessibility of the system parameters selected for observation, the prediction may not be very precise, which leads to false positive outcomes, or unforeseen events that are missed by the prediction algorithm. However, when errors or failures are predicted at a high degree of accuracy, avoidance or preventative actions may be applied. For example, event prediction may be used for proactive management in large-scale clusters \cite{Sahoo:2003}.  

\subsubsection{Restructure Pattern\\}
The occurrence of a fault, error, or failure event sometimes impacts a system in a way that affects the correctness of the interactions between sub-systems in an HPC environment, which causes further errors, or a failure of the system. The \texttt{Restructure Pattern}, a derivative of the \texttt{Reconfiguration} pattern, modifies the configuration between the interconnected sub-systems to isolate the specific sub-system affected by a fault, error or failure. The reconfiguration pattern alters the organization of the (sub-)systems to work around the affected (sub-)system, or it excludes the affected (sub-)system from interacting with the remaining (sub-)systems (i.e., the restructured system includes N-1 sub-systems). In either case, the pattern seeks to maintain (sub-)system functionality equivalent to that before the occurrence of the fault, error or failure event. 

The protection domain of the pattern spans the part of (sub-)system whose constituent sub-systems may be reconfigured. While the pattern seeks to restructure the sub-systems in an operating state that is functionally equivalent to the fault-free state, the pattern may result in the operation of the system in degraded condition, which incurs additional time overhead to the system. Existing solutions that restructure the system in response to an event include the ULFM extension to the MPI standard \cite{Bland:2013:IJHPCA}, which allows parallel applications to get notifications of process failures. ULFM provides a set of routines to revoke and restructure a MPI communicator that consists of the remaining active processes. Dynamic page retirement is another instantiation of the restructure pattern solution, in which pages that have an history of frequent memory errors are removed from the pool of available pages. \newpage

\subsubsection{Rejuvenation Pattern\\}
When a (sub-)system in an HPC environment behaves incorrectly on account of a fault, error or failure, the correctness of the full system may be compromised. The \texttt{Rejuvenation Pattern}, which is also a derivative of the \texttt{Reconfiguration} pattern, isolates the specific part of the (sub-)system affected by a fault, error or failure and restores it to an operating state that is free of any effects of the event.  Only the affected part of the system is rejuvenated to ensure correct operation of the system by the pattern. The pattern requires the system operation to be halted to identify the part of the system affected by the event. 

The protection domain of the pattern spans the part of system whose state may be rejuvenated. The rejuvenation is often a slow process that requires substantial additional overhead to identify the part of the system affected by the fault, error or failure, and to selectively reinitialize the system, in addition to overhead incurred due to any lost work. The rejuvenated system may not maintain the level of performance as before the occurrence of an event. Examples of rejuvenation include the Mini-Ckpts framework, which recovers fatal operating system crashes by rejuvenating only the kernel data structures, which are preserved in persistent memory, without affecting the HPC application state \cite{Fiala:2016}. Algorithm-based recovery methods for data corruptions in structures used in numerical analysis problems use interpolation of neighboring data values to rejuvenate data values in error state. Such methods have been demonstrated in the context of the Hartree−Fock algorithm used in computational chemistry codes \cite{vanDamm:2013:JCompChem}.

\subsubsection{Reinitialization Pattern\\}
The impact of a fault, error or failure may sometimes be irreversible such that the affected (sub-)system cannot be restored to a form that permits correct operation. The \texttt{Reinitialization Pattern}, also a derivative of the \texttt{Reconfiguration} pattern, simply restores the system to its initial state. This causes system operation to \textit{restart} with a pristine reset of state, which implicitly cleans up the effects of the fault, error or failure in the system. The pattern is applied in conditions in which the mitigation or recovery from the fault, error or failure event is deemed impossible, or excessively expensive in terms of overhead to performance. The pattern expects the fault, error or failure in the system to be detected; the pattern offers no implicit fault monitoring, prediction, or error/failure detection capability. The restoral of the system state to the initial state causes lost work, but guarantees the impact of the event is completely removed before service is resumed. Various cluster management software systems, such as the Cray Hardware Supervisory System (HSS) \cite{Cray:XE6Spec:2010}, enable malfunctioning nodes in the cluster to be reset. The HSS initiates a reboot sequence for a failing node without disrupting the remaining nodes in the system.

\subsubsection{Rollback Pattern\\}
Following an error or a failure event, the (sub-)systems in a HPC environment often lose all work performed until the occurrence of the event. The \texttt{Roll-back Pattern}, which derives from the \texttt{Checkpoint Recovery} architectural pattern, periodically captures the progress of the system and maintains these as system snapshots on a persistent storage system during the error/failure-free operation of the system. The rollback recovery is performed by restoring the system state based on the last known stable version of (sub-)system state. The solution provides rollback recovery, i.e., based on a temporal view of the system's progress, the system state restored during the error/failure recovery process is a previous error/failure-free state of the system. For a system that is deterministic, the pattern creates checkpoints of the system, which requires the capability to export the current (sub-)system state and import a new state during recovery. When the system design consists of several sub-systems, the pattern must coordinate the process of checkpointing. The instantiation of the pattern may apply the following coordination policies:
\begin{itemize}
\item \textit{Coordinated rollback recovery protocol}: The (sub-)systems coordinate the process of creating checkpoints, creating globally consistent checkpoint states, which simplify the recovery.
\item \textit{Uncoordinated roll-back recovery protocol}: The (sub-)systems each independently decide when to create their respective checkpoints. This approach has the potential to cause the full-system to propagate roll-backs to the initial system state to ensure that all dependencies are met (called the \textit{domino effect}).
\item \textit{Communication-based rollback recovery protocol}: The protocol enables each (sub-)system to create \textit{local} checkpoints, but periodically also enforces coordinated checkpoints between all (sub-)systems. Such a hybrid strategy helps avoid the \textit{domino effect}.
\end{itemize}

For systems with non-deterministic events, the pattern employs log-based protocols, which use a combination of checkpointing and logging of non-deterministic events in the (sub-)system. The log-based rollback recovery is based on piecewise deterministic assumption, in which the system identifies and records the nondeterministic events and information necessary (encoded in tuples called \textit{determinants}) to replay the event during recovery. The pattern may use the following logging protocols: 
\begin{itemize}
\item \textit{Pessimistic}: The protocol assumes that a failure occurs after a nondeterministic event in the system. Therefore, the determinant of each nondeterministic event is immediately logged to stable storage.
\item \textit{Optimistic}: The determinants are held in a volatile storage and written stable storage asynchronously. This protocol makes the assumption that the logging is completed before the occurrence of an error or failure. The error- or failure-free overhead of the optimistic approach is low.
\item \textit{Causal}: The protocol provides a balanced approach by avoiding immediate writing to stable storage (much like the optimistic protocol in order to reduce event free overhead), but each sub-system commits output independently (much like the pessimistic protocol in order to prevent creation of orphan sub-systems in the context of a multicomponent environment).
\end{itemize}

The protection domain for a \texttt{Rollback} pattern is determined by the extent of state captured during checkpoint operation and/or the number of system operations that can be recovered from the log of events. The time overhead introduced by the use of the pattern during error-free operation is correlated with the frequency of taking checkpoints. The rollback leads to loss of work due to the need to recover the system from a previous version of the system state. The amount of lost work is also correlated with the frequency of the checkpointing/logging. The worst-case scenario for recovery using this pattern is a roll-back to the initial state of the system. In the context of HPC systems, checkpoint and restart capabilities in the software layers, including various library-based and operating system-based solutions, enable recovery from process errors/failures and rollback of the applications. Well-known solutions that employ the rollback recovery pattern include the CoCheck checkpoint-restart for MPI \cite{Stellner:1996}, as well as BLCR \cite{BLCR:2002:LBNL} and SCR \cite{Mohror:2013:SCR}. Message logging protocols have been implemented in OpenMPI to support faster failure recovery \cite{Bouteiller:2010}.  

\subsubsection{Roll-forward Pattern\\}
When an error or failure event occurs in an HPC environment, a (sub-)system incurs loss of the work performed prior to the occurrence of the event. The \texttt{Roll-forward} pattern is a derivative of the \texttt{Checkpoint Recovery} pattern that avoids loss of work by using checkpoints to recover the (sub-)system to a stable state immediately before the error or failure event. Like the \texttt{Rollback} pattern, the solution entails the creation of snapshots of the system state and maintenance of these checkpoints on a stable storage system during the error- or failure-free operation of the system; log-based protocols use a combination of checkpointing and logging of non-deterministic events in the (sub-)system. However, the pattern uses the previously captured checkpointed state and/or logging information to recreate a stable version of the (sub-)system state identical to the one right before the error or failure occurred. This prevents the need for re-execution of all (sub-)system operations from the last stable checkpoint. The pattern must select checkpointing based on the policies similar to those used by the \texttt{Rollback} pattern: coordinated, uncoordinated, or communication-based.      

The pattern may use the following protocols for roll forward:
\begin{itemize}
\item \textit{Log-based protocols:} Based on the piecewise deterministic assumption, in which the (sub-)system uses the determinants to recreate state. The logging mechanisms may be based on \textit{pessimistic}, \textit{optimistic}, or \textit{causal} protocols.
\item \textit{Online recovery protocols:} Do not rely on event logging for roll forward of the (sub-)system; rather, they use inference methods to recreate state, or may permit the state to self-correct after restart.
\end{itemize}   

The protection domain of a \texttt{Roll-forward} pattern is determined by the extent of state captured during checkpoint operation and/or the number of system operations that can be recovered from the log of events. The pattern solution is not dependent on either the type of event, or the precise semantics of the error propagation; therefore, the design complexity in using this pattern in any HPC (sub-)system design in low. For the pattern to be effective in an HPC environment, the overhead to bring the system state to the most recent state before the error or failure must be less than or equal to the overhead of rollback recovery. In the context of HPC systems, software solutions typically implement roll-forward recovery using algorithm-specific knowledge. For example, Global View of Resilience (GVR) \cite{Chien:2016} uses versioning of distributed arrays supports, in which roll-forward recovery is based on application-specified mechanisms for each array structure. 

\subsubsection{N-modular Redundancy Pattern\\}
An error or failure of a (sub-)system in an HPC environment may cause loss in system capability or capacity, which prevents correct operation, or failure, of the full system. The \texttt{N-modular Redundancy Pattern}, which is a derivative of the \texttt{Redundancy} architectural pattern, remedies the effect of the error or failure by isolating the affected (sub-)system and compensating for its removal from the system design or configuration with a replica module. The solution entails creation of a group of N replicas of a (sub-)system. The replicated versions of a (sub-)system enables their use in various configurations to support errors or failures in one of the replicas, including fail-over, active comparison for error detection, or majority voting for detection and correction by excluding the replica whose outputs fall outside the majority. The pattern applies to a system with a modular design that has a well-defined scope and set of inputs and outputs. The scope of the pattern may be a sub-system in the HPC hardware or software architecture, or it may even encompass the complete system scope. Each of the N modules of the system exist simultaneously; the modules may be active at the same time (spatial replication), or may operate in succedent order (temporal replication), or the (sub-)system may activate the redundant modules on-demand. The protection domain of the pattern extends to the scope of the module that is replicated. Implementations of the MPI standard use these forms of redundancy for MPI messages, or even by replicating MPI process ranks; the MR-MPI \cite{Engelmann:2011}, rMPI \cite{Ferreira:2011:rMPI} and RedMPI \cite{Fiala:2012} are well-known MPI implementations using the n-modular redundancy approach. 

\subsubsection{Forward Error Correction Code Pattern\\}
When the state information of a (sub)-system is affected by an error, the incorrect state often leads to malfunctioning of the (sub-)system, which may lead to the failure of the full system. The \texttt{Forward Error Correction Code Pattern}, which is a derivative of the \texttt{Redundancy} architectural pattern, maintains redundant information about (sub-)system state. The pattern applies to a system whose state may be represented using a sequence of symbols. The solution consists of an encoder and a decoder module. In the simplest form, the encoder repeats each symbol that represents the (sub-)system state. The decoder module checks both instances of each state symbol. The general form of this pattern uses an encoder module that accepts \texttt{k} state information symbols and separately appends a set of \texttt{r} redundant symbols that are derived from the symbols representing (sub-)system state. The output of the encoder module is a (n, k) code, in which n = k+r. While the encoded redundant state information is a complex function of the original state, the encoder module does not modify the state information. The decoder module extracts the original state from the encoded state symbols. The availability of redundant state information enables recovery of system from corruption in symbols that represent the (sub-)system state by using the redundant information to reconstruct the original state information. 

The protection domain of the pattern extends to the scope of the (sub-)state that is encoded and decoded using the forward error correction code. The number of errors that are detectable and correctable is limited by the amount of redundant information contained in the error correction code. Since every operation that affects the system state requires encoding/decoding operations, the pattern introduces penalty in terms of time (increase in state information access latency), and space (increase in resources required to store state information) independent of whether an errors or failure occurs. There are various schemes that enable forward error correction in memory devices, storage systems, as well as in communication channels in HPC systems. Examples of forward error correction code (FEC) in HPC environments include parity bits, checksums, Hamming codes, hash function codes; more elaborate schemes such as systematic cyclic block codes include binary BCH, Reed-Solomon, Cyclic redundancy checks (CRC). The use of ECC in memory DIMMs is another well-known example of FEC for compensation of bit flip errors within the DRAM memory lines \cite{Moon:2005}. Algorithm-based methods use FEC schemes such as checksums to detect and correct errors in application data structures \cite{Huang:1984}.

\subsubsection{N-version Design Pattern\\}
When a design bug exists in a (sub-)system design or configuration, the resulting error or failure is often unavoidable. Therefore, the detection and mitigation of the impact of such errors or failures is critical. The \texttt{N-version Design Pattern}, which is a derivative of the \texttt{Design Diversity} pattern, applies distinct implementations of the same design specification created by different individuals or teams. The pattern applies N (N >= 2) independently implemented versions in a time or space redundant manner. The N versions of the (sub-)system are operated simultaneously, and a majority voting logic is used to compare the results produced by each design version. Due the low likelihood that different individuals or teams make identical errors in their respective implementations, the pattern enables compensating for errors or failures caused by a bug in any one implementation version. 

The pattern applies to a system that has a well-defined specification for which multiple implementation variants may be designed. The protection domain extends to the scope of the system that is described by the design specification. The extent to which each of the n versions are different affects the ability of the pattern to tolerate errors/failures in the system. The use of the n-version design pattern requires significant effort for design, implementation, testing and validation of the independent versions of a (sub-)system specification. Differences in the design may cause differences in timing in generating output values for comparison and majority voting; these differences incur overhead to the overall (sub-)system operation. 

\subsubsection{Recovery Block Pattern\\}
The errors and failures caused by design bugs prevent HPC (sub-)systems from operating in conformance with the (sub-)system specification. Yet, the application of the \texttt{N-version Design} pattern may be impractical in various contexts. The \texttt{Recovery Block Pattern}, which is also a derivative of the \texttt{Design Diversity} pattern, introduces an alternative implementation of the same design specification to perform detection and mitigation of errors. The pattern is a specialization of the \texttt{N-version Design} pattern since the solution also relies on multiple variants of a design that are functionally equivalent but designed independently. The recovery block is invoked when the result from the primary version of the system fails an acceptance test, which often indicates the presence of an error or failure. The instantiation of this pattern may sometimes include the function that performs the acceptance test.  
The consequence of applying the pattern in an HPC environment results in (sub-)system designs that consist of a module that implements the primary design and a module that serves as an exceptional case handler, i.e., the recovery block. There is also an adjudicator that applies an acceptance test to validate the results produced by the primary system. If the adjudicator does not accept the results of the primary system, it invokes the exception handler subsystem, which indicates that the primary system could not perform the requested service operation. The protection domain of the pattern extends to the scope of the primary system, i.e., the scope for which the recovery block is created. Examples of the recovery block pattern in HPC include the Containment Domains (CD) \cite{Chung:2011:SC} programming construct, which provides a recovery routine initiated upon detection of an error in the execution of the block of code encapsulated by the CD. This enables the CD to constrain the detection and correction of errors to the boundary of the domain.

%
%
\subsection{State Patterns}
\subsubsection{Static State Pattern\\}
The \texttt{Static} state pattern encapsulates all aspects of a system's state that is computed when the system is initialized, but is not modified during the system operation. The static state outlives the process that creates and initializes it. From the perspective of an HPC application, the static state includes program instructions and variable state that is computed upon application initialization. The correctness of the static state at all times is essential to the correct execution and outcome of a program. The invariant property of this state enables the use of a resilience behavioral pattern that can leverage this property to detect and recovery errors/failure of such state. For example, various algorithm-based fault tolerance methods leverage the property of invariance in the static state. These methods maintain replicas of the application variables in the static state pattern; recovery entails setting these variables to their default data values. A well-known application of this pattern is in the context of algorithm-based resilience techniques used in the design of iterative linear solvers. For the solution of a system of equations A.x = b, the static data structures such as the operand matrix A, the right-hand side vector B, or the preconditioner are computed once in the initialization phase of the application and are unchanged after. Errors in these structures are recovered using maintaining checksums \cite{Huang:1984}.

\subsubsection{Dynamic State Pattern\\}
The \texttt{Dynamic State} pattern encapsulates the state that changes as the system makes forward progress. In an HPC application, the pattern refers to all aspects of the program state that changes as an application program executes. The dynamic state includes the data variables that are modified by the algorithm, as well as the control-flow variables that enable forward progress of the system. The \textit{dynamic} feature of this state pattern implies that any faults or errors in such state amounts to lost work. Separating the dynamic state enables the identification of the appropriate behavioral resilience patterns to detect and correct errors in such state. Due to the transitory nature of the variables in the dynamic state patterns, the behavioral patterns often require preservation of the state pattern, or repetition of operations from a known stable point to recreate a version of the variables in the state pattern that are free from the effects of any errors. The most well-known method for protecting dynamic state is using checkpointing-based roll-back recovery methods \cite{Elnozahy:2002}. 

\subsubsection{Environment State Pattern\\}
The \texttt{Environment State Pattern} encapsulates the system state that plays a supporting role in the operation of the system. The pattern defines the scope of the system state that provides a common set of services in support of the primary function of the system. The environment also facilitates and coordinates the operation of various sub-systems in a system. In general, HPC systems navigate complexity through the definition of abstractions that hide the details of specific functions behind well-defined interfaces. When executing an HPC application, the overall system state may be partitioned into the aspects that are related to the application program state and those that provide access to the system resources and services that enable the application to fulfill its function. The pattern enables the resilience behavior of the environment state to be reasoned about separately from the resilience behavior of the primary system state, i.e., an HPC application. The separation of the environment state enables designers to instantiate behavioral patterns that are independent of the design of the algorithms of HPC applications. Any changes in the environment due an error or failure event directly affects the application program operating within the environment. 
While an application program does not normally have complete control over its environment, it may exert partial control to affect the environment through well-defined interfaces. The \texttt{Environment} state pattern defines the scope of the state that support resource sharing, coordination and communication between the various (sub-)systems. In a typical HPC system stack, the environment state pattern includes productivity tools and libraries, the runtime system, the operating system, file systems, communication libraries, etc. For example, operating-system based resilience mechanisms are independent of the resilience features of the application program and solely focus on the correctness of the data structures within the kernel. Mini-Ckpts is a known example of a framework that emphasizes the recovery of the OS environment by preserving kernel structures in persistent memory \cite{Fiala:2016}. Similarly, the ULFM MPI provides recovery of the communication environment from the failure of processes by reconstructing the MPI communicator by creating consensus among the remaining set of processes \cite{Bland:2013:IJHPCA}.

\subsubsection{Stateless Pattern\\}
The \texttt{Stateless} pattern enables the definition of resilience solutions that are independent of system state. Since every resilience solution consists of at least a state and behavior pattern, the \texttt{Stateless} pattern
provides the construct of \textit{null} state in order to create solutions that have a well-defined notion of behavior, but don't define a scope for the behavior. From the perspective of an HPC application, the definition of the \texttt{Stateless} pattern permits the definition of the scope of operations that perform detection or recovery without explicitly specifying the variable state of the program that is affected by the operations. The solutions that are based on a \texttt{Stateless} pattern may include: (i) applications that consist of predominantly memory load operations that rarely contain state-modifying memory and I/O operations; these applications typically perform reduction operations over large number of data elements, and (ii) applications that yield imperfect results since their algorithms are based on approximation and iterative refinement, or use noisy input data to begin with. 
The stateless pattern is utilized together with behavioral resilience patterns whose actions do not necessitate modifying any particular aspect of the system state during the detection or recovery. However, the resilience solution that uses a stateless pattern must select and instantiate a behavioral pattern that is capable of dealing with any additional side-effects due to the inclusion of the stateless pattern. The use of the \textit{transaction} model to provide resilient behavior is an example of the \texttt{Stateless} pattern. Transactions support execution of a sequence of operations that may complete as a unit, or fail; the notion of partial execution is not supported. For example, in the Relax framework, the idempotence property guarantees that any region can be freely re-executed, even after partial execution, and still produce the same result. Relax supports language-level constructs as well as compiler-based techniques that enable the definition of idempotent regions of execution; the recovery of such regions are stateless \cite{deKruijf:2010:ISCA}.

\section{Building Resilience Solutions using Design Patterns}
\label{sec:Solutions-Classification}

\begin{figure*}[tp]
\centering
\includegraphics[width=125mm, height=50mm]
                {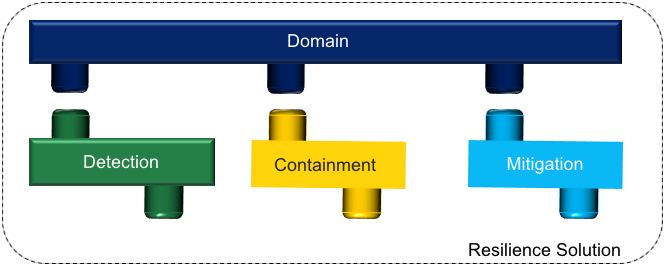}
\caption{Elements of a resilience solution for HPC systems and applications}
\label{Fig:PatternSolutionElements}
\end{figure*}
\subsection{Components of Resilience Solutions}
Each pattern in the resilience design pattern catalog presents a solution to a specific problem in detecting, containing or mitigating a fault, error or failure event. However, ensuring that an HPC application executes to result in a correct solution despite the occurrence of the events in the systems requires that a resilience solution be constructed using multiple such patterns that are organized in a well-defined system of patterns. The artifacts of a design process that uses design patterns are complete resilience solutions that confront a specific type of event and provide detection, containment and mitigation capabilities over a well-defined protection domain. Therefore, the first step in the design of a solution is the selection of patterns for each of these capabilities. Therefore, a complete solution consists of at least one state pattern (defining scope of the protection domain) and one or more behavioral patterns (supporting a combination of detection, containment and mitigation solutions). These key constituents of a complete solution are shown in Figure \ref{Fig:PatternSolutionElements}. 

The pattern descriptions allow for instantiating each pattern in the catalog at any layer of the system stack. The individual patterns that make up a complete solution can be implemented across layers the system stack. The architecture of a HPC system consists of various types of processor, memory, storage and networking components, and its software stack is a complex multicomponent environment consisting of communication and threading libraries, productivity software and tools, including numerical libraries, runtime systems, profiling tools, etc. To construct resilience solutions for the hardware and software components requires methodically selecting resilience patterns that may be conveniently incorporated into the design of these components. The coordination between the resilience patterns, particularly when implemented across layers of abstraction, requires well-defined activation and response interfaces for each pattern.  

\subsection{Design Spaces}
\begin{figure}[tp]
\centering
\includegraphics[width=0.25\textwidth]
                {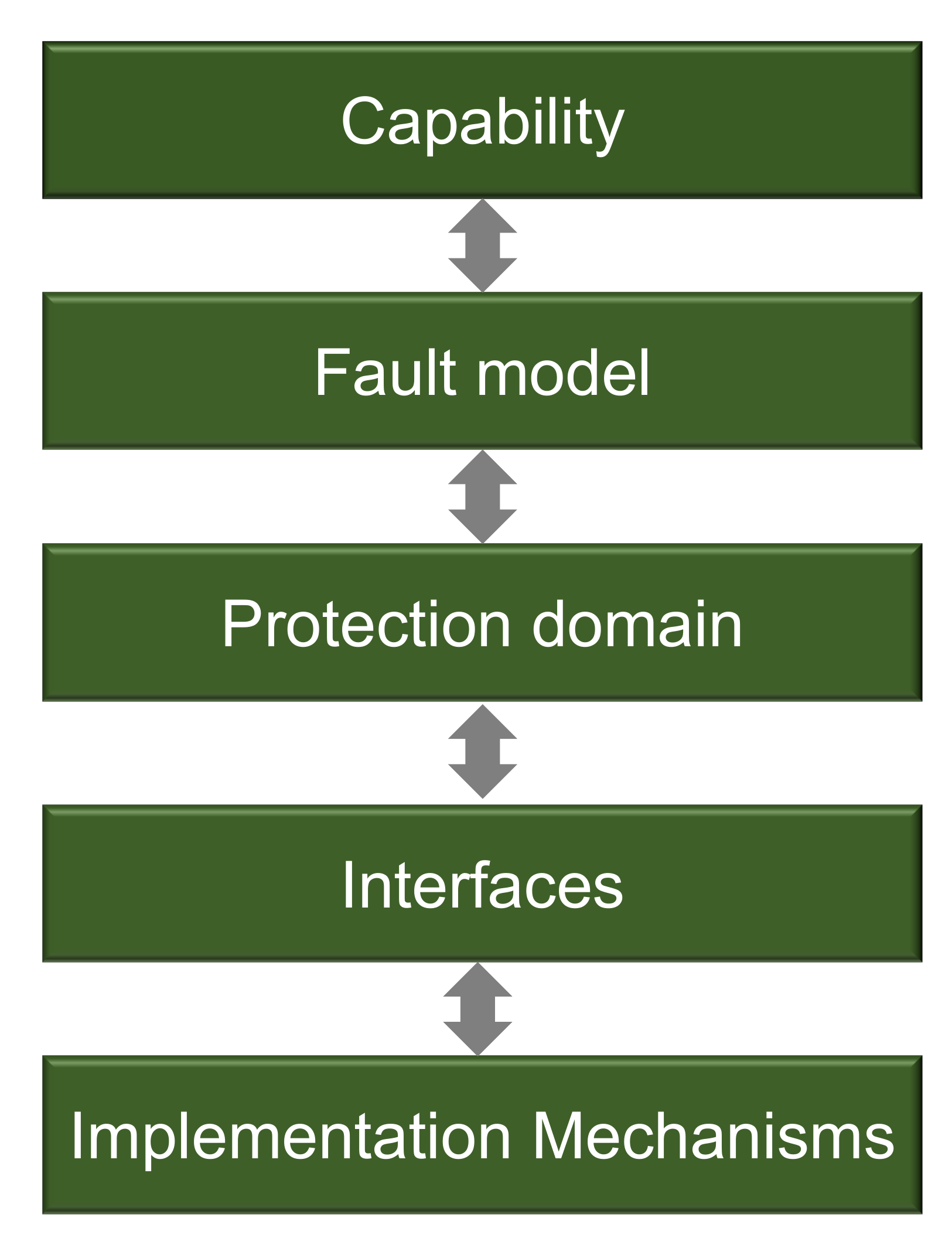}
\caption{Design Spaces for construction of resilience solutions using
         patterns}
\label{Fig:PatternDesignSpaces}
\end{figure}

For hardware and software designers to make practical use these patterns in the development of resilient versions of their designs, a set of guidelines are necessary to combine the patterns and refine their interactions. The hierarchical classification scheme articulates only certain aspects of the pattern selection and integration process by categorizing the patterns based on the type of event they handle and the core technique employed. However, the selection of patterns solely on the basis of their detection, containment and mitigation capabilities leaves much to skills of the designer in terms of finalizing the design and the implementation of the component or system. To build practical resilience solutions various other factors must be considered, including the layer of abstraction for their implementation, scalability of the solution, portability to other architectures, dependencies on any hardware/software features, flexibility to adapt the solution to accelerated fault rates, capability to handle other types of fault and error events, the performance and performance overheads.   

To enable a systematic assessment of the suitability of a resilience pattern to a specific context and to integrate patterns into composite solutions, we develop a design framework. The framework enables the creation of an initial outline of the resilience solution that identifies the strategy patterns and the captures the dimensions and capabilities solution resulting from the composition of the patterns. The framework is based on {\em design spaces} that are arranged in a hierarchy. Each design space progressively refines the relationships between the patterns and optimizes the overall solution, which allows for a structured approach for constructing customized designs. By navigating over the design spaces, the framework enables the designer to approach the various issues that must be addressed in the process of developing practical resilience solutions. The framework, which is illustrated in Figure~\ref{Fig:PatternDesignSpaces}, consists of the following design spaces: 

\begin{itemize}
\item \textbf{Capability}:
      This design space is concerned with identifying the patterns that support capabilities for the detection, containment, mitigation of a specific type of fault, errors or failure event. Based on the system context, this design space also considers the organization of the overall structure of the solution.  
\item \textbf{Fault model}:
      By identifying the root causes of fault and understanding the impact and propagation through the system enables deciding the architecture patterns. The design space emphasizes the selection of architecture patterns and the distribution of responsibility among the chosen patterns.   
\item \textbf{Protection domain}:
      This design space concentrates on the definition of the protection domain by deciding the state patterns and their composition. This enables a clear encapsulation of the system scope over which the resilience patterns operate.
\item \textbf{Interfaces}:
      The identification and implementation of the activation and response interfaces for behavioral patterns affect the propagation of fault/error/failure event information. Within this design space, the layer of abstraction appropriate for the instantiation of the pattern, as well as the performance and power overheads are considered. The design space explores various implementation constructs that facilitate the coordination between the various patterns, particularly across system layers.  
\item \textbf{Implementation mechanisms}:
      This design space is concerned with low-level implementation details of how patterns are embedded within a hardware structure, or in software code. It considers the constraints imposed by specific features of hardware, execution or programming models, software environment and how the various pattern implementations coordinate their behavior in this context. 

\end{itemize}

The design spaces represent the most important aspects of a resilience solution that a designer must contemplate in order to create effective and efficient resilience solutions. As a designer navigates through these design spaces, they are able to develop a clearer understanding of the solution profile and the general constraints, which enables them to select the appropriate patterns from the catalog and decide on implementation alternatives. The use of resilience patterns in the context of the framework provided by the design spaces enables HPC system designers, users and application developers to evaluate the feasibility and effectiveness of novel resilience techniques, as well as analyze and evaluate existing solutions. They provide a structured flow to the design process the design spaces articulate the critical decision points in the design of a resilience solution, providing guidelines for the selection of the appropriate patterns based on the requirements of protection and the cost of using specific patterns. 

Designers may use various approaches to navigate the design spaces, including a strictly top-down approach, in which the design is driven by the event type and model that a system must be protected against, and the implementation of the system is adapted to enable the system to survive the different ways in which the event may impact the reliability of the system. Alternatively, in a bottom-up approach, the resilience capability must be woven into the existing hardware or software component designs and interfaces, and additional components are included to enhance the protection coverage, or to handle specific fault model behaviors. Often, designers may be required to take a hybrid approach, in which the design spaces are revisited in an effort to refine a design, to optimize the features of a solution, and to enable designers to overcome constraints imposed by any hardware or software system features.

\section{Case Studies}

This section explores use cases for the application of resilience design patterns to the systematic design and analysis of resilience solutions. We use the pattern-based approach for understanding existing solutions with the view to adapt the solution to future generations of HPC systems as well as for exploration and assessment of novel cross-layered solutions. The case studies describe the pattern-based design process for different fault models on a notional architecture and software environment of a HPC system.   

\subsection{Checkpoint and Rollback Solution for Process Failures}
\begin{figure*}[t]
\centering
\includegraphics[width=0.80\textwidth,height=70mm]{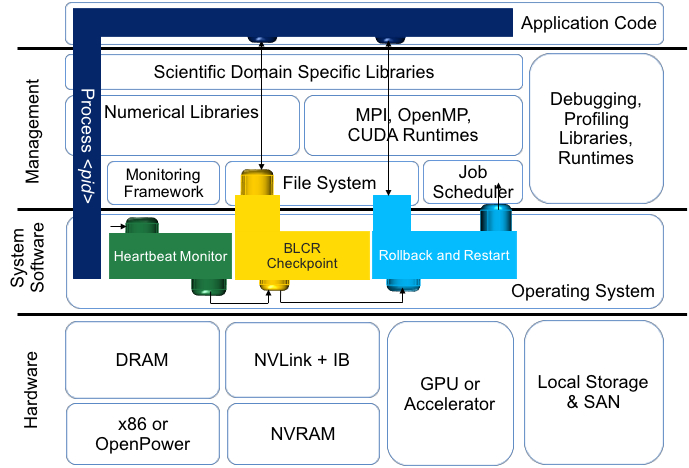}
\caption{Case Study: Checkpoint \& Restart-based Recovery}
\label{Fig:PatternCaseStudy-CR}
\end{figure*}

For this case study, we aim to develop a resilience solution that enables an HPC application to survive process failures. In an HPC environment, the diagnosis of the precise root cause of these failures is difficult due to the lack of sufficient hardware-level debugging information. For designing a purely software-based solution, the fault model is a process crash or hang whose cause is unknown. This type of failure results from the presence of a fault in the processor or memory that activates, which causes an error in the form of an illegal instruction, or an invalid address in the program state. When the program execution encounters the address in the program state that is in error state, the process may crash or hang. 

Checkpoint and restart (C/R) solutions are the often used to support resilience to process failures in HPC systems. We reexamine this well-known software-based solution using the structured pattern-based approach to analyze composition of the constituent patterns needed to design this solution. Such analysis will be useful for adapting C/R solutions to future systems and evaluate their performance characteristics. The goal of a complete C/R solution is to recover a failed process such that the application may resume from an error-free state. This requires that the solution capture the image, or snapshot, of a running process and preserves it for later recovery. For parallel applications, the C/R framework's coordination protocols produce a global snapshot of the application by combining the state of all the processes in the parallel application. Since most parallel applications using the message passing interface (MPI) define a MPI process to be a POSIX process, the protection domain of the solution must cover the complete POSIX process state. Therefore, we fuse the \texttt{Persistent} and \texttt{Dynamic} and \texttt{Environment} state patterns, which extends the domain of our system-level checkpointing solution to the entire memory associated with a process. In a Linux-based environment, the protection domain covers the total virtual address space of a Linux process.

For the detection of a process failure, we require instantiation of the \texttt{Fault Treatment} strategy pattern. Specifically, our solution requires a \texttt{Fault Diagnosis} architecture pattern to discover the location of the failure and the type of event, which is enabled by a \texttt{Monitoring} structural pattern. The instantiation of the \texttt{Monitoring} pattern is a kernel-level heartbeat monitor, which is deployed in the system to detect whether the process is alive.

For the selection of a recovery pattern, there are key two considerations: (i) the frequency of node failures; and (ii) the performance and resource overhead of applying the pattern. The space overhead incurred by instantiating a \texttt{Compensation} strategy pattern for recovery is substantial due to the need to replicate the protection domain. For systems that experience process failures infrequently, the use of a compensation-based solution proves prohibitively expensive. Therefore, for the failure recovery we select the \texttt{Recovery} strategy pattern. The \texttt{Checkpoint-Recovery} architectural pattern is appropriate since Linux provides the capability for a running process to be interrupted and its context to be written to disk. Also, the process state is deterministic and defined by the state of the program counter and the registers; therefore, the \texttt{Roll-back} structure pattern is suitable for implementation at the operating system level. With the selection of this pattern protection domain of the failure to be limited to a single process context, which implicitly defines the containment pattern. The implementation of the recovery pattern requires a disk storage system, to which the checkpoint, i.e., the process state captured during failure-free operation is exported. The performance overhead of these patterns during failure-free operation and the recovery time are dependent on bandwidth available between memory and the disk system.  

The implementation of the patterns, which is illustrated in Figure \ref{Fig:PatternCaseStudy-CR}, is implemented using the Berkley Lab's Checkpoint/Restart (BLCR) \cite{BLCR:2002:LBNL} framework. Since BLCR does not provide a failure detection mechanism, the \texttt{Monitoring} pattern is implemented by a kernel-level module that uses heartbeat monitoring to check for process liveness. BLCR provides a completely transparent checkpoint of the process, which saves the current state of a Linux process. The framework uses a coarse-grain locking mechanism to momentarily interrupt the execution of all the threads of the process, giving them a global view of its current state. The entire state is saved, including the CPU registers, the virtual memory map as well as the function call stack. From the perspective of an application programmer, the checkpoint routine returns with a different error code, to let the caller know if this function call returns from a successful checkpoint or from a successful restart. The \texttt{Roll-back} pattern handles recovery after the detection of a process failure by restoring the context file set from the stable storage, and recreating the process on the same hardware, with the same software environment. BLCR also provides an API for applications programmers to manage pattern behavior through hooks that allow the application to block off code sections where checkpoints are not permitted. These hooks also give applications a chance to respond to checkpoint/requests and take appropriate action, which provides an application programmer with explicit control over the pattern's activation and response interfaces.  

\subsection{Proactive Process Migration for Failure Avoidance}
In HPC environments, various fault indicators indicate the imminence of error or failure events. The goal of this case study is to design and implement a proactive resilience solution using the structured design pattern-based approach. In contrast to a reactive solution that seeks to recover from an error or a failure event after the fact, a proactive solution identifies faults in a system and seeks to remedy the anomaly or defect to prevent their activation to result in errors or failures. This analysis of this solution is intended to identify the patterns that must instantiated for a proactive design approach, and to articulate the protection domain of the solution. 

\begin{figure*}[t]
\centering
\includegraphics[width=0.80\textwidth, height=70mm]{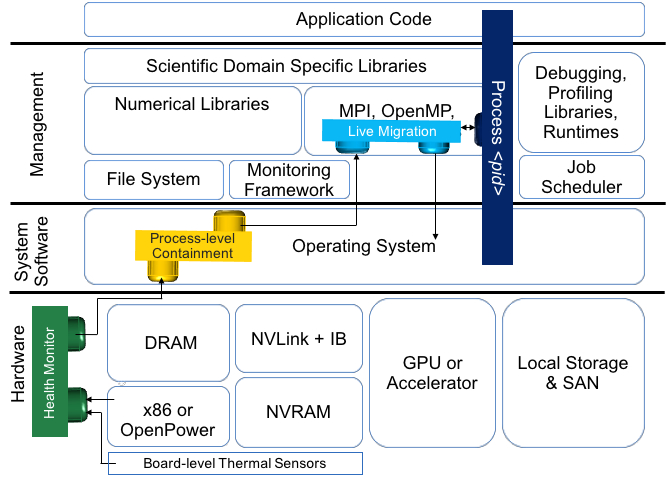}
\caption{Case Study: Proactive Process Migration}
\label{Fig:PatternCaseStudy-Migration}
\end{figure*}

The key to designing a proactive strategy is the identification of fault indicators that can sufficiently predict the activation of an error or failure. The fault model for this case study is a defect in the system that has the potential to result in an error or failure. We consider faults that are known to cause errors, which result in application crashes. Using design patterns, we seek to develop a software-based solution that can preemptively migrate parts of an application away from system resources that are about to fail. In a HPC system, the failure of a compute node causes termination of the application processes running on that node. Since the presence of a fault does not impact the correctness of an application program until it activates, the solution supports proactive failure avoidance from the application's perspective. We select the protection domain by fusing the \texttt{Persistent} and \texttt{Dynamic} and \texttt{Environment} state patterns. Much like the C/R solution, the protection domain covered by these patterns includes the complete POSIX process state in a Linux environment. The ultimate objective of the solution is to preemptively migrate the application processes from compute nodes where a failure is likely to cause them to crash to another node in the system.      

To anticipate the occurrence of a failure, the solution must observe critical indicators that will predict the likelihood of a failure. We apply the \texttt{Fault Treatment} strategy pattern, which is instantiated as a \texttt{Fault Diagnosis} pattern in every node of the HPC system. This pattern is instantiated as a \texttt{Prediction} structural pattern, which enables estimating the possibility of an imminent error or failure event. Its activation interface reads health monitoring data for the various components in each compute node and its response interface signals the possibility of a node failure. The prediction pattern creates a control feedback-loop such that a mitigation pattern can take preventive action to avoid failure of the processes running on the node. Since the solution addresses faults in the computes nodes, it requires the instantiation of another \texttt{Fault Treatment} pattern for mitigation rather than a \texttt{Recovery} strategy pattern. For this solution, we assume that the number of nodes allocated for an application run are determined during startup and are fixed for the lifetime of the application run. If the application uses all nodes in the allocation at initialization and leaves no spare nodes, the inclusion of a \texttt{Compensation} strategy pattern is not a suitable alternative. The \texttt{Reconfiguration} architectural pattern is applied, which is instantiated in the form of a \texttt{Restructure} structural pattern that isolates a failing node and migrates the application processes to an alternative compute node in the system. The containment is implemented by a kernel level module provides containment for the fault by identifying the process that is executing on the node which the \texttt{Prediction} pattern has assessed vulnerable due to a specific set of changes in operating conditions of the node.

The overall structure of the pattern-based design is illustrated in Figure \ref{Fig:PatternCaseStudy-Migration}. The implementation of the \texttt{Prediction} pattern is realized as a per-node health monitoring mechanism that uses various platform-level indicators in the system. It uses platform data available through the Intelligent Platform Management Interface (IPMI) interface, which relies on the baseboard management controller (BMC) to collect sensors readings for health monitoring, including the data on temperature, fan speed, and voltage. The response interface of the pattern notifies the scheduler when the sensors indicate deterioration of a node's health. Since the behavior of the \texttt{Recovery} strategy pattern used by this solution entails performing a live migration of a POSIX process in the context of the MPI execution environment, the implementation of the \texttt{Restructure} pattern is realized within the system's job scheduler. The pattern identifies healthy nodes in the system as potential destinations for the process migration. Once a destination node has been identified, the pattern initiates the migration of the process from source to destination node. It is imperative the entire context of a process be migrated when the presence of a fault is inferred on a compute node. Therefore, the migration entails transfer of the process image, which occurs by a page-by-page copy of the address space. The implementation then synchronizes all the MPI processes to a consistent state, after which the in-flight data in the MPI communication channels is drained.  Once all the MPI processes reach a consistent global state, the remaining dirty pages, which includes the registers, signal information, pid, files, etc. to the destination node. Once the mapping of the processes to nodes in the system has been restructured, the communication channels and the previously saved in-flight messages are restored. The migrated processes resume execution on the destination node. The implementation of the patterns in this solution ensure the transparency of the proactive migration to the HPC application. 

\subsection{Cross-layer Hardware/Software Solution for Soft Error Resilience}
\begin{figure*}[t]
\centering
\includegraphics[width=0.80\textwidth, height=70mm]{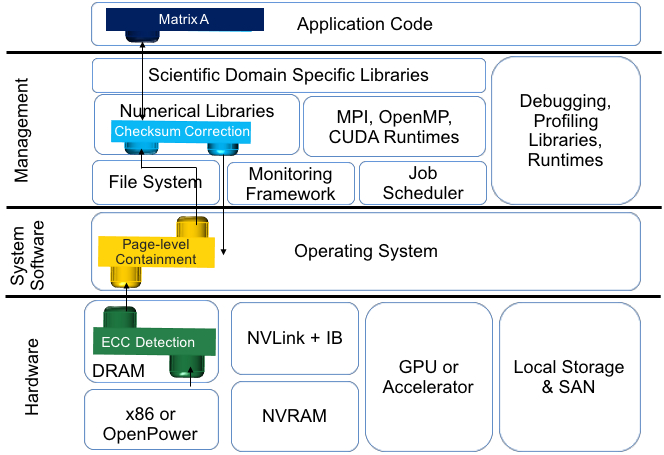}
\caption{Case Study: Cross-Layer Design using Algorithm-based Fault Tolerance to complement Hardware-level Error Correction Codes}
\label{Fig:PatternCaseStudy-CrossLayer}
\end{figure*}

In this case study, we use design patterns as building blocks to explore novel resilience solutions that leverage capabilities from various layers of the system stack. By navigating the design spaces of the resilience design pattern framework, we can evaluate the effectiveness of instantiating a detection, containment or mitigation pattern at a specific level in the system stack and systematically construct a cross-layer resilience solution that connects patterns from multiple layers. The structured approach supported by the framework also enables refining the cross-layered solution. The aim of this case study is to develop a solution that provides soft error detection and correction for HPC application data structures. The fault model that we consider is transient errors in memory structures that cause multiple bit flips in the application's data or control variables, which may result in outcomes ranging from incorrect results to fatal program crashes.   

The DRAM memory chips used in HPC systems use error correcting codes (ECC) to detect and correct bit flip errors. Similarly, algorithm-based fault tolerance techniques are available that maintain checksums for data structures to detect and correct data value errors at the application level. However, the lack of formal methods to combine these solutions often precludes cross-layer hardware-software designs that cooperative protect the application data. Our proposed solution is designed to support transient error resilience for a scientific application that uses an iterative linear solver method. In general, these methods   solve a system of linear equations represented as A.x = B, where x is the solution vector, A is the operand matrix and b is a known vector. The iterative algorithm begins with an initial approximation of the solution x, and refines this solution until the residual norm is below a certain error bound. Therefore, the matrix A and vector b are scoped within \texttt{Static} state patterns, the solution vector x in a \texttt{Dynamic} state pattern, and the remaining variable state is contained within an \texttt{Environment} pattern. While the solution vector is often tolerant to perturbations due to the iterative nature of the algorithm, any transient errors within the scope of the two \texttt{Static} state patterns affects the correctness of the solver. Therefore, we define the protection domain of our cross-layer solution to include only these static patterns.  

For achieving error detection and correction in digital data, the general approach is to add redundant information to discover errors and reconstruct the original data. This approach fits the \texttt{Compensation} strategy pattern, which may be instantiated in the form of a \texttt{Forward Error Correction} pattern. For the detection of the transient errors, we assume that this pattern is implemented in the form of ECC in the DRAM modules, which supports single-bit error correction and double-bit error detection. Therefore, the instantiation of this structural pattern handles both detection and mitigation for single-bit errors. Double-bit errors result in an ECC violation on the memory line, which is asynchronously communicated by the \texttt{Forward Error Correction} pattern to the operating system via its response interface by raising a machine check exception. For the containment of the double-bit error, we deploy a \texttt{Fault Treatment} pattern in the operating system, since the OS views the double-bit corruption as a fault. Since the pattern must discover whether the double-bit corruption maps to the protection domain specified by the state patterns, it is instantiated as a \texttt{Fault Diagnosis} pattern, specifically as a \texttt{Monitoring} structural pattern. For recovery of variable state scoped by the \texttt{Static} state pattern, the solution instantiates the \texttt{Compensation} strategy pattern. It uses the \texttt{Redundancy} architecture pattern and structures the solution based on the \texttt{Forward Error Correction} pattern.  

The instantiation of the patterns across the system stack is illustrated in Figure \ref{Fig:PatternCaseStudy-CrossLayer}. The \texttt{Monitoring} pattern for containment is implemented as a kernel-level module that maps the physical address to the virtual address space to discover whether the fault may be contained within the \texttt{Static} state pattern. The pattern's response interface treats the presence of the fault in the state pattern as an application error and notifies the numerical library. When the error is outside the scope of the \texttt{Static} state pattern, the response interfaces indicates to the kernel module that the error is unrecoverable, which results in the OS killing the application. Besides the \texttt{Forward Error Correction} pattern in ECC for single-bit error recovery, another instance of this pattern type is implemented in the numerical library to handle double-bit errors. The implementation maintains a set of checksums for the matrix A and vector b. The checksums enable the identification of the element of the matrix affected by the error, and substitution of that element with a correct value using the remaining uncorrupted elements in the row/column and the checksum values. The instantiation of the \texttt{Forward Error Correction} pattern at the application library level provides context about the significance of the error to the overall application, and is able to employ an algorithm-specific fault tolerance detection and correction method, which is more cost effective for double-bit error mitigation than system-level bulk reliability provided by hardware-level solution such as an enhanced ECC that supports double-bit correction. Therefore, the cooperation between patterns across system layers supports a flexible memory protection mechanism to single and double-bit memory errors, which allows the application to resume operation towards completion rather than experience a fatal crash with higher performance and energy efficiency.

\section{Related Work}

The original concept of design patterns was developed in the context of civil architecture and engineering problems where patterns were defined with the goal of identifying and cataloging solutions to recurrent problems and solutions in the building and planning of neighborhoods, towns and cities, as well as in the construction of individual rooms and buildings \cite{Alexander:1977}. In the domain of software engineering, patterns were introduced in an effort to bring discipline to the art of programming and create reusable designs. The intent of software design patterns isn't to provide a finished design that may be transformed directly into code; rather, these patterns are used to systematize the software development process by using proven paradigms and methodologies in software engineering practice \cite{Buschmann:1996}. With the use of design patterns, there is sufficient flexibility for software developers to adapt their implementation to accommodate any constraints, or issues that may be unique to specific programming paradigms, or the target platform for the software. Related to software design patterns, the concept of algorithmic skeletons was introduced \cite{Cole:1991} and further refined \cite{Cole:2004}. In the context of object-oriented (OO) programming, design patterns provide a catalog of methods for defining class interfaces and inheritance hierarchies, and establish key relationships among the classes \cite{Gamma:1995}. In many OO systems, reusable patterns of class relationships and interactions between objects are used to create flexible, elegant, and ultimately reusable software design. Pattern systems have also been developed for cataloging concurrent and networked object-oriented environments \cite{Schmidt:2000}, resource management \cite{Kircher:2004}, and distributed systems \cite{Buschmann:2007}.

In the pursuit of quality and scalable parallel software, patterns for programming paradigms were developed \cite{Mattson:2004} as well as a pattern language, called Our Pattern Language (OPL) \cite{Mattson:OPL:2009}. These describe the computation and communication patterns in various parallel algorithms and therefore useful for designing and implementing scalable parallel applications. For engineering parallel applications for shared-memory many-core processors, parallel programming patterns simplify the process of expressing parallelism using a number of programming interfaces such as OpenMP, OpenCL, Cilk Plus, ArBB, Thread Building Blocks (TBB) \cite{McCool:2012}. Patterns also support the implementation of parallel algorithms that automatically avoid unsafe race conditions and deadlocks \cite{McCool:2010}.

Design patterns have been discovered in a variety of other domains and used to codify the best-known solutions, which include patterns for natural language processing \cite{Talton:2012}, user interface design \cite{Borchers:2001}, web design \cite{Duyne:2002}, visualization \cite{Heer:2006}, and software security \cite{Dougherty:2009}. Patterns have also been defined for enterprise applications that involve data processing in support or automation of business processes \cite{Fowler:2002} in order to bring structure to the construction of enterprise application architectures. In each of these domains of design, the patterns capture the essence of solutions in a succinct form such that they may be easily applied to other contexts.

Previous efforts to develop design patterns for fault tolerance have defined a number of patterns for error detection, recovery and mitigation. These patterns are developed based on well-known fault tolerance solutions that are used in mission-critical systems such as telecommunication systems and space programs \cite{Hanmer:2007}, distributed systems \cite{Saridakis:2002} and enterprise data processing systems \cite{Friedrichsen:2012}. The fault tolerant version of the Common Object Request Broker Architecture (CORBA) \cite{Natarajan:2000} applies patterns in the design of the middleware to improve the performance of a range of fault tolerance strategies that provide applications with capabilities for rapid recovery from service failures, including request-retry, redirection, active and passive replication. While the capabilities of some of the patterns in these domains overlap with the resilience patterns described in this document, they solve problems that are significantly different from those encountered in HPC environments in terms of the system architectures, the software stack, and the nature of the applications. The patterns in this document specifically address the challenges for maintaining resilient operation for HPC systems and their applications.

\section{Summary}

In this paper, we introduce the concept of resilience design patterns, which support a systematic approach to designing and implementing resilience solutions. The structured approach to the design of HPC resilience solutions is useful to reduce the complexity of the design process, and is particularly relevant for the future generations of extreme-scale parallel systems and their applications. The resilience design patterns are based on well-known and well-understood solutions that have been applied in HPC systems and provide solutions to specific problems encountered in the management of resilience. The patterns presented in this document support detection, containment, masking and recovery capabilities. The resilience patterns may be used by designers as reusable templates when building and refining new resilience solutions and for reengineering existing solutions for future generations of HPC systems. The paper also presents a classification scheme that organizes the resilience patterns in a layered hierarchy in order to expose the relationships between the various patterns in the catalog and their capabilities. The hierarchical organization of the patterns enables system hardware and software architects to approach the solution at an abstract level, while individual component designers and software developers may restrict their work to the level that directly impacts their portion of the solution. We have also developed a design framework to simplify the composition of design patterns into complete resilience solutions. The framework is useful for navigating the various design challenges and constraints encountered by designers and enables the creation of flexible and portable resilience solutions. The resilience patterns and the pattern-oriented framework also facilitates the exploration of alternative solutions, the refinement and optimization of solutions, and the investigation of the effectiveness and efficiency of solutions. This structured approach aims to address the resilience challenge for extreme-scale HPC systems through a systematic design of solutions with an emphasis on optimizing the trade-off, at design time or runtime, between the key system design factors: performance, resilience, and power consumption.

\ack{This material is based upon work supported by the U.S. Department of Energy, Office of Science, Office of Advanced Scientific Computing Research, program manager Lucy Nowell, under contract number DE-AC05-00OR22725.}
\bibliography{references}

\end{document}